\pdfoutput=1 

\documentclass[a4paper,aps,prl,onecolumn,notitlepage,superscriptaddress,showpacs,amsmath,amssymb]{revtex4-1}

\usepackage{geometry}
\usepackage{graphicx}
\usepackage{wasysym}
\usepackage{xcolor}

\begin{document}
	
	\title{Disentangling the roles of roughness, friction and adhesion in discontinuous shear thickening by means of thermo-responsive particles}
	\author{Chiao-Peng Hsu}
	\affiliation{Laboratory for Soft Materials and Interfaces, Department of Materials, ETH Zurich, Zurich, Switzerland.}
	\affiliation{Laboratory for Surface Science and Technology, Department of Materials, ETH Zurich, Zurich, Switzerland.}
	\author{Joydeb Mandal}
	\affiliation{Laboratory for Surface Science and Technology, Department of Materials, ETH Zurich, Zurich, Switzerland.}
	\author{Shivaprakash N. Ramakrishna}
	\affiliation{Laboratory for Surface Science and Technology, Department of Materials, ETH Zurich, Zurich, Switzerland.}
	\author{Nicholas D. Spencer}
	\affiliation{Laboratory for Surface Science and Technology, Department of Materials, ETH Zurich, Zurich, Switzerland.}
	\author{Lucio Isa}
	\email{lucio.isa@mat.ethz.ch}
	\affiliation{Laboratory for Soft Materials and Interfaces, Department of Materials, ETH Zurich, Zurich, Switzerland.}

	\begin{abstract}
		Dense suspensions of colloidal or granular particles can display pronounced non-Newtonian behaviour, such as discontinuous shear thickening (DST) and shear jamming (SJ). The essential contribution of particle surface roughness and adhesive forces confirms that stress-activated contacts can play a key role in these phenomena. By employing a system of microparticles coated by responsive polymers that allow friction, adhesion, and surface roughness to be selectively and independently tuned as a function of temperature, we offer a way to disentangle these contributions.  We find that DST occurs at lower shear rates  when friction and adhesion between particles are enhanced at high temperatures. Additionally, the temperature-responsive polymers provide lubricity at low temperatures that can mask surface roughness. The link between single-particle properties and macroscopic rheology is elucidated via lateral force microscopy, which reveals the nature of rheologically relevant contact conditions. \textit{In situ} temperature tuning during shear allows contact conditions to be modified, and DST to be switched on and off on demand. These findings strengthen our understanding of the microscopic parameters affecting DST and identify new routes for the design of smart, non-Newtonian fluids.
	\end{abstract}
	\maketitle
	
	
	Shear thickening (ST) is a generic phenomenon that occurs when the shear stress $\sigma$ increases faster than linearly with the shear rate $\dot{\gamma}$, so that the viscosity $ \eta \equiv \sigma/\dot{\gamma}$ effectively increases with shear rate. It can be observed in a broad range of materials, but nowhere is it more prominent than in dense suspensions of solid particles \cite{Brown_Shear_2014}. This phenomenon may take the severe form known as discontinuous shear thickening (DST) \cite{Wagner_Shear_2009, Brown_Generality_2010, Seto_Discontinous_2013, Fernandez_Microscopic_2013, Brown_Shear_2014, Wyart_Discontinuous_2014} , where the suspension's viscosity increases by orders of magnitudes at a critical shear rate, or, in the most extreme cases, the suspension may even solidify under shear -- an occurrence known as shear jamming (SJ) \cite{Cates_Jamming_1998, Waitukaitis_Impact-activated_2012, Peters_Direct_2016}. Both instances can lead to failures in high-shear processes, but can also be exploited for applications, e.g. in granulation or impact-absorption. Recent studies have shown that interparticle contacts play a crucial role in DST, triggered by a change in interaction between particle surfaces from hydrodynamic lubrication to boundary lubrication at high shear \cite{Seto_Discontinous_2013, Fernandez_Microscopic_2013, Wyart_Discontinuous_2014, Lin_Hydrodynamic_2015, Royer_Rheological_2016, Clavaud_Revealing_2017, Morris_Lubricated_2018, Singh_From_2019}. While fluid films allow suspended particles to easily slide past each other at low shear, beyond a critical shear stress the hydrodynamic lubrication films between particles break down, resulting in particles that are effectively in asperity-asperity contact and thus can engage in frictional interactions due to boundary lubrication. This transition indicates that the surface morphology and surface chemistry of the particles can have a striking influence on the macroscopic flow behaviour. Based on this knowledge, an engineering-tribology approach can be utilized to design the thickening of suspensions. 
	
	From an experimental standpoint, interparticle friction coefficients can be effectively controlled by engineering surface chemistry \cite{Fernandez_Microscopic_2013} or by modifying surface roughness \cite{Lootens_Dilatant_2005, Hsiao_Rheological_2017, Hsu_Roughness-dependent_2018}. Moreover, short-range adhesive forces, e.g., as experimentally introduced by hydrogen bonding \cite{James_Interparticle_2018, James_Tuning_2019}, can also be used to modify the properties of interparticle contacts and strongly affect the rheology. However, regulating the interparticle tribology while the suspensions are being sheared as a means to offer external control on the flow properties has not yet been addressed. Finally, disentangling the different contributions that link the tribology of contacts to the rheology of the suspension remains an elusive task with important consequences for materials design and fundamental understanding alike. 
	
\begin{figure}
		\centering
		\includegraphics[width=1\textwidth]{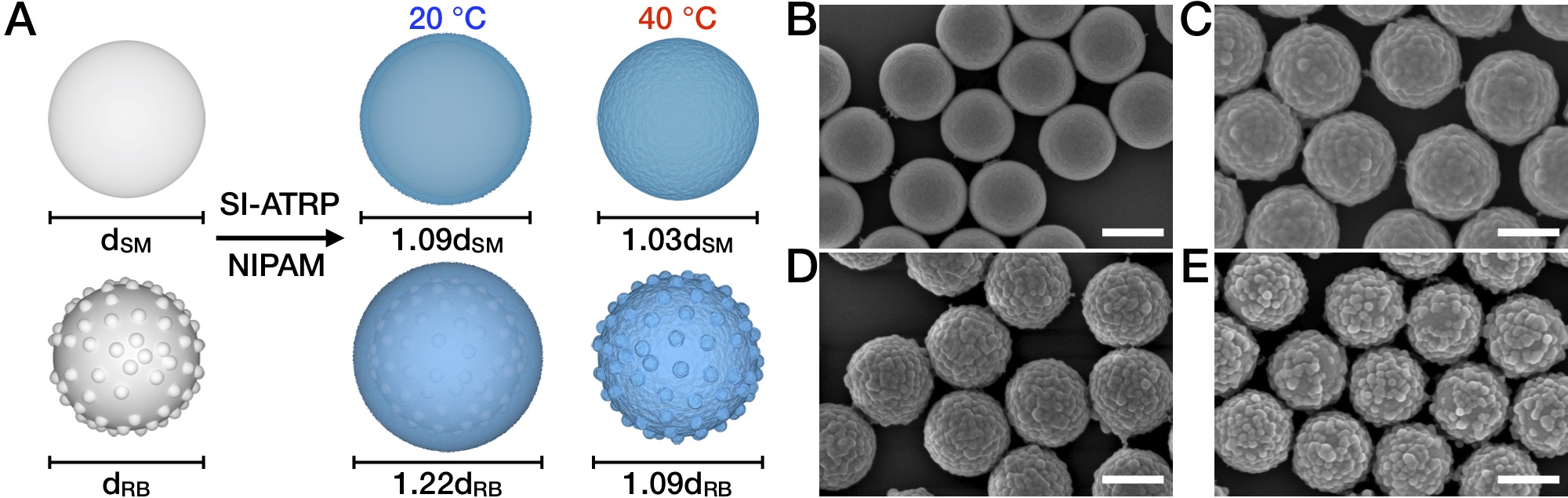}
		\caption{\textbf{PNIPAM-grafted smooth and rough silica particles.} \textbf{a}, Schematics of the PNIPAM-grafted smooth (SM) and rough (RB) particles at 20 $^\circ$C and 40 $^\circ$C. d$_{\text{SM}}$ and d$_{\text{RB}}$ are the diameters of smooth and rough particles before polymerization, respectively. \textbf{b--e}, Scanning electron microscope images of four PNIPAM-grafted particle systems: SM\_PNIPAM (\textbf{b}), RB\_0.36\_PNIPAM (\textbf{c}), RB\_0.46\_PNIPAM (\textbf{d}), RB\_0.52\_PNIPAM (\textbf{e}). Scale bars indicate 500 nm.}
		\label{Fig1}
	\end{figure}

	To address this, we examine model silica colloids of varying surface roughness coated with thermo-responsive polymer brushes of poly(N-isopropylacrylamide) (PNIPAM). Comparing the nanotribology and rheology of these model colloids allows us to test the roles of friction, adhesion, and surface roughness independently and to tune them during shear. We synthesize raspberry-like silica particles with controllable roughness \cite{Hsu_Roughness-dependent_2018} and employ surface-initiated atom transfer radical polymerization (SI-ATRP) \cite{Matyjaszewski_Polymers_1999, Mandal_Tuning_2019} to graft PNIPAM brushes from the particles' surfaces (see Methods and Figure S1). A distinctive feature of PNIPAM brushes is that they undergo a swelling-deswelling transition in water across a lower critical solution temperature (LCST) of 30--33 $^\circ$C \cite{Halperin_Poly_2015}. Figure \ref{Fig1} displays the different PNIPAM-grafted particles employed in this study. Smooth silica PNIPAM-grafted particles are named “SM\_PNIPAM” and rough silica PNIPAM-grafted  particles are named “RB\_$h/d$\_PNIPAM”, where $h/d$ is the value of the dimensionless roughness parameter of the silica surface measured before polymerization ($h$ is the average asperity height and $d$ is the average inter-asperity spacing obtained by atomic-force-microscopy imaging of the particles' surfaces) \cite{Hsu_Roughness-dependent_2018}. The dehydration of PNIPAM brushes across the LCST results in different effective particle sizes as a function of temperature. In particular, we designed the swollen thickness of the PNIPAM brushes, $h_{\text{PNIPAM}}$, to ensure that the underlying roughness is masked at 20 $^\circ$C and revealed at 40 $^\circ$C, as shown in Figure \ref{Fig1}a. Using SI-ATRP offers exquisite control over the brush thickness by selecting the precise solvent ratio, catalyst ratio, and polymerization time, hence enabling unique tailoring of the physico-chemical properties of model, brush-stabilized particles. The properties of the PNIPAM-grafted particles are summarized in Table \ref{Tab1}.
	
	\begin{table}[htbp]
		\caption{\textbf{Details of the PNIPAM-grafted particles.} $R_\text{h}(T)$ and $R_{\text{h}_0}(T)$ are the particle hydrodynamic radii after and before polymerization, respectively, at the given temperature. $h_{\text{PNIPAM}}(T)$ is the PNIPAM thickness, for which $h_{\text{PNIPAM}}(T) = R_\text{h}(T)$ - $R_{\text{h}_0}(T)$. $m_{\text{PNIPAM}}/m_{\text{SiO}_2}$ is the mass fraction of PNIPAM brushes and silica particles.}
		\centering
		\begin{tabular}{ccccc}		
			\hline 
			&SM\_PNIPAM &RB\_0.52\_PNIPAM &RB\_0.46\_PNIPAM &RB\_0.36\_PNIPAM  \\ 
			\hline   
			$R_\text{h}(20 ^{\circ}C)/R_{\text{h}_0}(20 ^{\circ}C)$ &1.09 $\pm$ 0.02 &1.22 $\pm$ 0.02 &1.23 $\pm$ 0.03 &1.22 $\pm$ 0.03  \\
			$h_{\text{PNIPAM}}(20 ^{\circ}C)$ (nm) &31 $\pm$ 3 &80 $\pm$ 5 &85 $\pm$ 6 &84 $\pm$ 6 \\
			$R_\text{h}(40 ^{\circ}C)/R_{\text{h}_0}(40 ^{\circ}C)$ &1.03 $\pm$ 0.01 &1.08 $\pm$ 0.02 &1.09 $\pm$ 0.02 &1.08 $\pm$ 0.02  \\
			$h_{\text{PNIPAM}}(40 ^{\circ}C)$ (nm) &12 $\pm$ 2 &31 $\pm$ 3 &33 $\pm$ 3 &35 $\pm$ 4 \\
			$m_{\text{PNIPAM}}/m_{\text{SiO}_2}$ &0.11 &0.12 &0.13 &0.13  \\
			\hline
		\end{tabular}
		\label{Tab1}
	\end{table}
	
	\begin{figure*}
		\centering
		\includegraphics[width=1\textwidth]{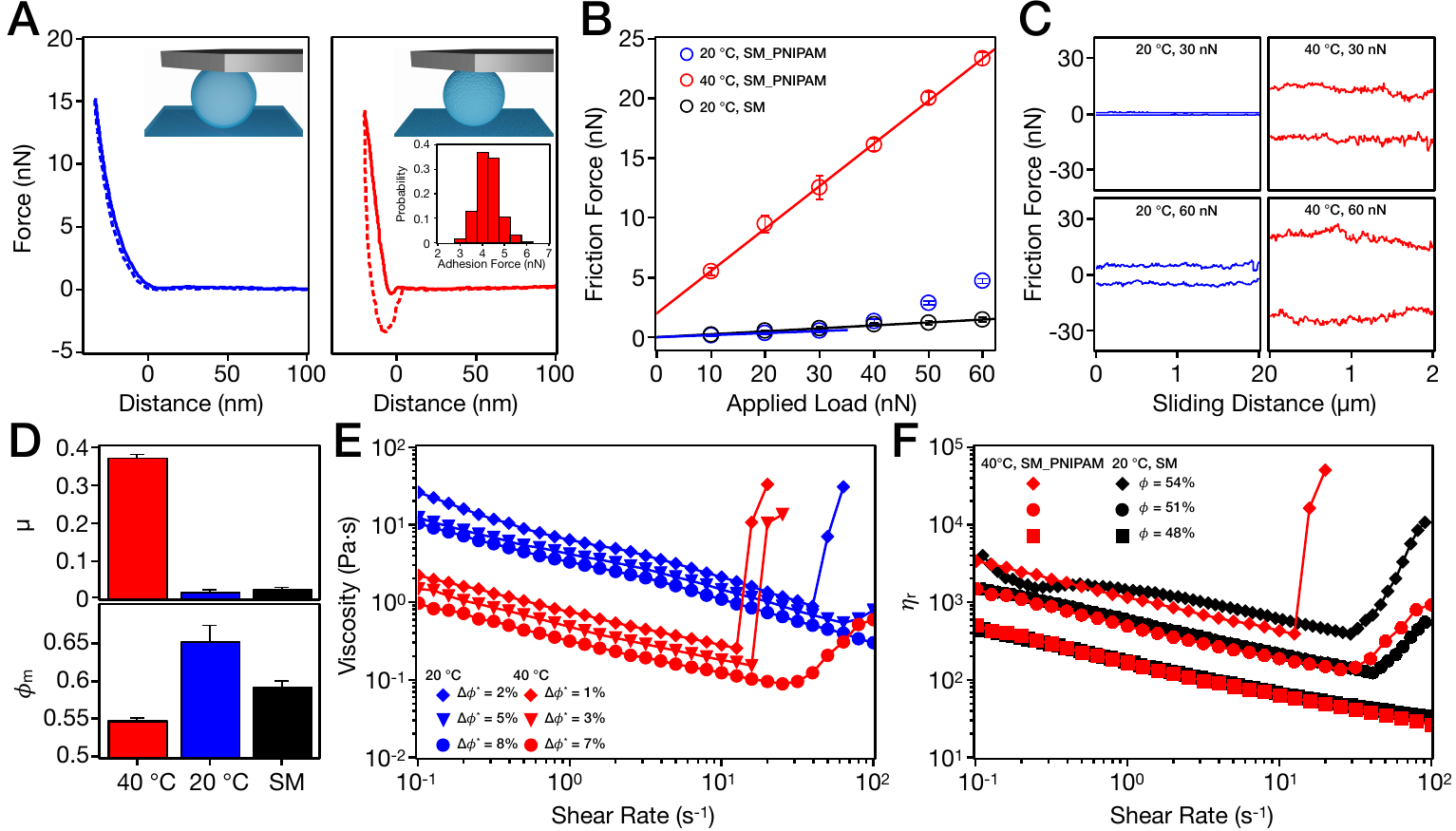}
		\caption{\textbf{Tribology and rheology experiments on PNIPAM-grafted smooth particles.} \textbf{a}, Approach (solid line) and retraction (dashed line) of the force--distance curves of SM\_PNIPAM system at 20 $^\circ$C (left) and 40 $^\circ$C (right). The top two insets show the schematics of a SM\_PNIPAM probe on a countersurface at 20 $^\circ$C (left) and 40 $^\circ$C (right). The bottom left inset shows the adhesion forces measured at 40 $^\circ$C. The bin width is 0.5 nN. \textbf{b}, Friction-force-vs-applied-load measurements of SM\_PNIPAM at 20 $^\circ$C (blue) and 40 $^\circ$C (red), and SM (black). Error bars represent the standard deviations across the scan area (51 data points). \textbf{c}, Friction loops of the SM\_PNIPAM system at 30 nN (top); 60 nN (bottom), and at 20 $^\circ$C (left, blue); 40 $^\circ$C (right, red). \textbf{d}, Friction coefficients $\mu$ (top) and frictional packing fraction $\phi_m$ (bottom) of the SM\_PNIPAM system at 40 $^\circ$C (red) and 20 $^\circ$C (blue), and SM system (black). Error bars in $\mu$ represent the uncertainties in the calculated slope based on Equation \ref{Eqn1}. $\mu$ of SM\_PNIPAM at 20 $^\circ$C was calculated from 10 to 30 nN. Error bars in $\phi_m$ represent the standard deviation from four repeat measurements. \textbf{e}, Rheological flow curves of SM\_PNIPAM in rescaled packing fractions $\Delta \phi^{*}$ at 20 $^\circ$C (blue) and 40 $^\circ$C (red). \textbf{f}, Normalized flow curves of SM\_PNIPAM at 40 $^\circ$C (red) and SM at 20 $^\circ$C (black) in $\phi$ = 48, 51, and 54\%. The relative viscosity $\eta_r$ was calculated based on $\eta_r = \eta_{\text{measured}}/\eta_{\text{water}}(T)$. The results for the SM system are adapted from previous work \cite{Hsu_Roughness-dependent_2018}.}
		\label{Fig2}
	\end{figure*}
	
	\begin{figure}
		\centering
		\includegraphics[width=1\textwidth]{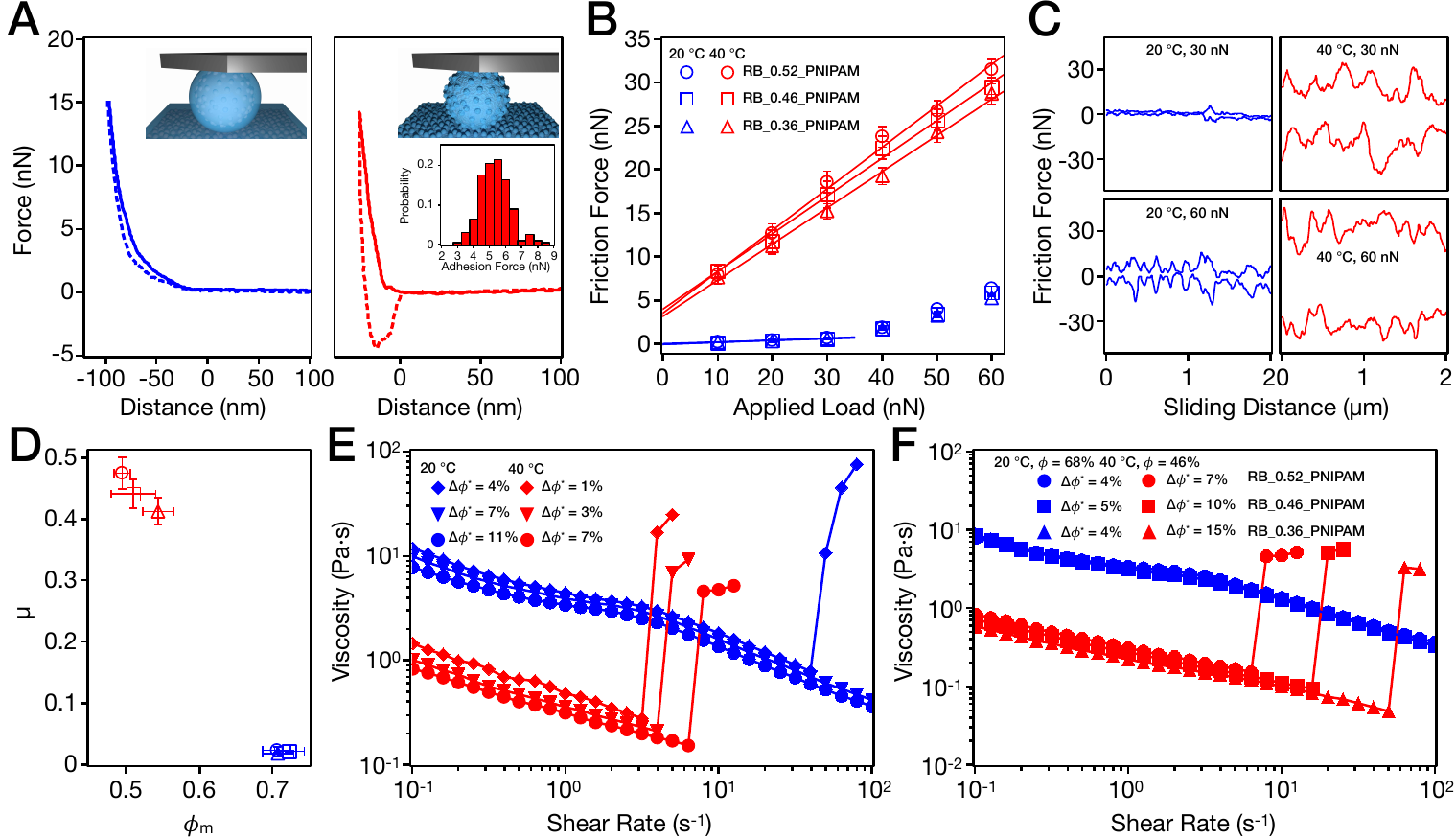}
		\caption{\textbf{Tribology and rheology experiments on PNIPAM-grafted rough particles.} \textbf{a}, Approach (solid line) and retraction (dashed line) of the force--distance curves of RB\_0.52\_PNIPAM system at 20 $^\circ$C (left) and 40 $^\circ$C (right). The top two insets show the schematics of a RB\_0.52\_PNIPAM probe on a countersurface at 20 $^\circ$C (left) and 40 $^\circ$C (right). The bottom inset shows the adhesion forces measured at 40 $^\circ$C. The bin width is 0.5 nN.\textbf{b}, Friction-force-vs-applied-load measurements of RB\_0.52\_PNIPAM ($\bigcirc$), RB\_0.46\_PNIPAM ($\square$), and RB\_0.36\_PNIPAM ($\triangle$) at 20 $^\circ$C (blue) and 40 $^\circ$C (red). Error bars represent the standard deviations across the scan area (51 data points). \textbf{c}, Friction loops of the RB\_0.52\_PNIPAM system at 30 nN (top); 60 nN (bottom), and at 20 $^\circ$C (left, blue); 40 $^\circ$C (right, red). \textbf{d}, $\mu$ as a function of $\phi_m$ of the three PNIPAM-grafted rough systems (same legends as \textbf{b}) at 20 $^\circ$C (blue) and 40 $^\circ$C (red). $\mu$ at 20 $^\circ$C was calculated from 10 to 30 nN. Error bars in $\mu$ and $\phi_m$ represent the uncertainties in the calculated slope based on Equation \ref{Eqn1} and the standard deviation from four repeat measurements, respectively. \textbf{e}, Rheological flow curves of RB\_0.52\_PNIPAM in various values of $\phi$ at 20 $^\circ$C (blue) and 40 $^\circ$C (red). \textbf{f}, Rheological flow curves of the three PNIPAM-grafted rough systems with $\phi$ = 68\% at 20 $^\circ$C (blue) and $\phi$ = 46\% at 40 $^\circ$C (red).}
		\label{Fig3}
	\end{figure}
	
	\begin{figure}
		\centering
		\includegraphics[width=0.6\textwidth]{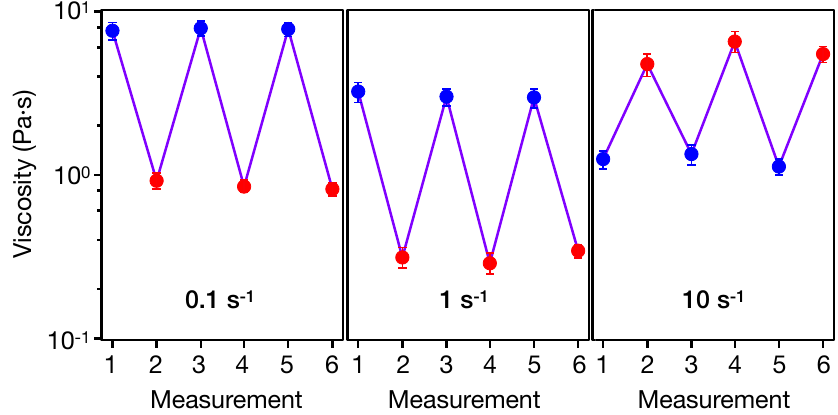}
		\caption{\textbf{Thermo-switchable shear-thickening system.} Viscosity measurements at shear s of 0.1 s$^{-1}$ (left), 1 s$^{-1}$ (middle), and 10 s$^{-1}$ (right) of RB\_0.52\_PNIPAM at 20 $^\circ$C; $\phi$ = 68 \% (blue) and 40 $^\circ$C; $\phi$ = 46 \% (red). }
		\label{Fig4}
	\end{figure}
	
	The solubility transition of PNIPAM brushes not only affects their thickness but also has a strong impact on their surface interactions. In order to characterize this effect at the single-contact level, we measure the tribological properties of PNIPAM-grafted particles by colloidal-probe atomic force microscopy (AFM). In particular, we provide controlled contact conditions by attaching individual PNIPAM-coated particles of a given roughness onto an AFM cantilever and sliding them against a flat, PNIPAM-coated counter-surface of the same underlying roughness at two temperatures -- above (40 $^\circ$C) and below the LCST (20 $^\circ$C), respectively -- measuring friction as a function of applied load. The substrates are produced by a process analogous to the synthesis of the polymer-coated colloids, to provide representative, realistic counter-surfaces mimicking interparticle contacts in the sheared suspensions (see Methods and Figure S2). 
	
	Crossing the LCST changes both the adhesion and the friction between PNIPAM-coated surfaces \cite{Yu_Stretching_2015}. To decouple these contributions from particle topography, we first examine the case of smooth particles (Figure \ref{Fig2}). We start by characterizing the adhesion force between PNIPAM-grafted particles at different temperatures by measuring the pull-off force in a force-versus-distance curve (see Methods). Representative force--distance curves at 20 $^\circ$C and 40 $^\circ$C are shown in Figure \ref{Fig2}a. The retraction curve below LCST shows no adhesion force. Conversely, above the LCST, a small jump-to-contact is observed during approach and the retraction curve displays marked adhesion with a pull-off force of 4.2 $\pm$ 0.9 nN. This adhesion is due to the hydrophobic interaction between the collapsed PNIPAM brushes \cite{Chang_Switchable_2007, Jones_Variable_2002}. We then measure the friction by means of lateral force microscopy (see Methods). The friction forces above the LCST are $\sim$20 times higher than below the LCST, as shown in Figure \ref{Fig2}b and a non-zero friction force at zero load is found, in agreement with the presence of adhesion. Moreover, at 20 $^\circ$C, the SM\_PNIPAM particles and smooth silica particles without the polymer (SM) display almost identical low friction behavior at loads below 30 nN. At higher loads, the SM\_PNIPAM particles start to show higher friction forces than SM particles. Further details of the surface interactions were obtained by measurement of the friction force versus sliding distance in the so-called "friction loops” (Figure \ref{Fig2}c). The narrow loop for SM\_PNIPAM at 30nN is indicative of smooth sliding and low friction. Upon increasing the load to 60nN, the friction loop opens up, indicating increased dissipation, which has been ascribed to the entanglement of PNIPAM brushes at a higher contact pressure \cite{Divandari_Topology_2017}. The friction loops at 40 $^\circ$C always show higher dissipation and irregular sliding, as a consequence of adhesion. 
	
	In the presence of adhesion force, the relation between friction force $F_{\text{friction}}$ and applied load $L$ in Figure \ref{Fig2}b allows a friction coefficient to be extracted using  a modified version of Amontons' Law \cite{Derjaguin_Molecular_1934}:
	\begin{equation}
	F_{\text{friction}} = \mu\cdot(L_{0}+L) = F_{0}+\mu\cdot L,
	\label{Eqn1}
	\end{equation}
	where a constant internal load $L_{0}$ is added to the applied load $L$, in order to account for the intermolecular adhesive forces. $ F_{0} $ represents the friction force at zero applied load for adhesive surfaces. The friction coefficient $\mu$ is then defined as the slope $dF_{\text{friction}}/dL = \mu$, which we plot in Figure \ref{Fig2}d. 
	
	Our previous works have shown that the interparticle friction coefficient is directly correlated to the high-shear maximum packing fraction obtained in centrifugation experiments $\phi_m$ (see Methods) \cite{Fernandez_Microscopic_2013, Hsu_Roughness-dependent_2018, Luo_One_2020}. We also observe here that a system having higher $\mu$ has a lower $\phi_m$ (Figure \ref{Fig2}d). However, we notice that, although the $\mu$ values of SM\_PNIPAM and SM are fairly close at 20 $^\circ$C, there is $\sim$10\% difference in their $\phi_m$ values. This discrepancy is due to the fact that SM\_PNIPAM has a soft and compressible shell, resulting in a denser packing during sedimentation and highlights that the detailed relationship between $\mu$ and $\phi_m$ may be system specific. 
	
	The distinct adhesion and friction behaviors of SM\_PNIPAM at 20 $^\circ$C and 40 $^\circ$C also lead to distinct shear-rheology responses as a function of temperature. In Figure \ref{Fig2}e, we examine the flow curves of three dense suspensions of SM\_PNIPAM. Because of the swelling and dehydration of the PNIPAM brushes, each suspension has a different effective $\phi$ at different temperature, e.g., $\phi$ = 64\% at 20 $^\circ$C and $\phi$ = 54\% at 40 $^\circ$C for the same suspension. However, qualitative differences beyond a rescaling of the volume fraction emerge, and these are clearly visible if we rescale the volume fraction so that $\Delta \phi^{*}=(\phi_m - \phi)/\phi_m$ . SM\_PNIPAM colloids at 20 $^\circ$C start displaying continuous shear thickening (CST) for $\Delta \phi^{*}$ = 5\% and exhibit DST at $\Delta \phi^{*}$ = 2\% -- very close to their measured $\phi_m$ of 65.2\% at 20 $^\circ$C. At 40 $^\circ$C, adhesion and higher friction switch on, which shift the onset of ST to lower shear rates and enhance its magnitude. The same suspension exhibits CST at 40 $^\circ$C ($\Delta \phi^{*}$ = 7\%), while showing shear thinning at 20 $^\circ$C ($\Delta \phi^{*}$ = 8\%). At higher solids loading, the same suspension transitions between CST at 20 $^\circ$C ($\Delta \phi^{*}$ = 5\%) to DST at 40 $^\circ$C ($\Delta \phi^{*}$ = 3\%). For the densest suspension where dilatant DST is observed at both temperatures (positive N$_1$, see Figure S3), the onset rate and stress (Figure S3) occur at significantly lower values at  40 $^\circ$C.  
	
	To isolate the effect of volume fraction from that of contact properties, it is worth comparing the flow curves of low-frictional and non-adhesive SM colloids without PNIPAM at 20 $^\circ$C with the high-frictional and adhesive SM\_PNIPAM colloids at 40 $^\circ$C at the same $\phi$ values (Figure \ref{Fig2}f). Strikingly, for the shear-thinning and CST regimes with $\phi$ up to 51\%, both suspensions behave identically, even if the contact properties are radically different, confirming that the hydrodynamic volume fraction is the dominating parameter in describing the rheology. Conversely, a marked difference is observed concerning the DST behavior. At $\phi$ = 54\%, SM\_PNIPAM and SM display similar low-shear rheology, but at high shear rate the former exhibits DST and the later shows CST. This finding reinforces the view that contact properties strongly affect DST over and above hydrodynamic effects. 
	
	Moving from SM to RB particles, additional features emerge. The RB\_PNIPAM systems also display switchable adhesion across the LCST (Figure \ref{Fig3}a and Figure S4), with an adhesion force of 5.3 $\pm$ 1.3 nN between RB\_0.52\_PNIPAM and its countersurface at 40 $^\circ$C. This value is not very different from that measured for the SM\_PNIPAM, confirming the analogy in the chemical origin of the adhesive bonds. As opposed to the smooth particles' case, the precise control of brush thickness allows us to switch on and off the underlying surface roughness of the RB\_PNIPAM particles at different temperatures. This switching has a striking influence on the friction behaviour of the RB\_PNIPAM systems, as shown in Figure \ref{Fig3}b and \ref{Fig3}c. At 20 $^\circ$C, all three RB\_PNIPAM systems behave like SM\_PNIPAM and SM, exhibiting identical and low friction below 30 nN. This fact indicates that the swollen PNIPAM shell effectively screens the roughness at low loads, as borne out by the flat, narrow friction loop for RB\_0.52\_PNIPAM at 30 nN. Upon increasing load, i.e. to 60 nN, stick--slip appears in the friction loop due to the interlocking of asperities, which are now revealed as the polymer brush is highly compressed, even below the LCST (Figure \ref{Fig3}c). Conversely, at 40 $^\circ$C, the RB\_PNIPAM systems show high, roughness-dependent friction, with higher friction forces measured for particles with higher underlying roughness (Figure \ref{Fig3}b). Essentially, two factors contribute to the increase of friction above the LCST: the adhesion between the PNIPAM brushes and the unveiling of roughness. Above the LCST, the surface of the RB\_PNIPAM particles turns sticky and rough because of the collapse of PNIPAM brushes, allowing asperities to interlock within sticky contacts. At 40 $^\circ$C, we thus observe wide friction loops with stick--slip events at all loads, as shown in Figure \ref{Fig3}c.  
	
	As a consequence of the differences in the frictional behavior at the different temperatures for the three RB\_PINPAM particles studied, a clear relation between $\phi_m$ and $\mu$ emerges, as shown in Figure \ref{Fig3}d. Confirming previous observations, we see that the higher the effective interparticle friction coefficient, the lower the value of $\phi_m$ \cite{Fernandez_Microscopic_2013, Hsu_Roughness-dependent_2018}. Because the effect of topography emerges only above the LCST, the differences between the three systems emerge only at higher temperatures. Conversely, our particles show almost identical friction coefficient and $\phi_m$ values below the LCST, again supporting the conclusion that swollen PNIPAM shells effectively screen roughness. As already introduced for the smooth particles, the temperature-dependent tribology of the rough colloids strongly affects their rheology at high packing fractions, as shown for dense suspensions of RB\_0.52\_PNIPAM in Figure \ref{Fig3}e. At 20 $^\circ$C, the suspensions do not display any appreciable CST for $\Delta \phi^{*}$ up to 7\%, but directly show DST at 4\%. This behavior is analogous to the previously reported ST response of uncoated, rough silica spheres \cite{Hsu_Roughness-dependent_2018} and stems from the interlocking of asperities when a high shear stress breaks through the PNIPAM lubrication layer, as clearly shown by the presence of stick--slip events in the friction loop at high load (Figure \ref{Fig3}c). The same suspensions display a distinctively different response at 40 $^\circ$C. All three suspensions show dilatant DST (positive N$_1$, see Figure S5), and the onset rate for DST shifts to lower shear rates and shear stresses (Figure S5) with increasing $\phi$. DST is even observed for values of $\Delta \phi^{*}$, for which the suspension below the LCST only shows shear thinning, i.e., $\Delta \phi^{*}$ = 7\%. The strongly enhanced DST above the LCST is generated by the sticky, rough particle surface, where in addition to asperity interlocking, adhesion between the collapsed PNIPAM brushes causes a significant increase in the effective friction. 
	
	Finally, we compare the shear rheology of the three RB\_PNIPAM systems at the same $\phi$ as shown in Figure \ref{Fig3}f. The flow curves obtained below the LCST with $\phi$ = 68\% are all identical and show shear thinning, implying that with no ST the rheology is defined by the packing fraction alone. Above the LCST, surface differences emerge only during ST, whose onset becomes roughness dependent. Below the onset of DST, all suspensions, which have also the same $\phi$ at 40 $^\circ$C, show the same viscosity, but the different roughness causes the breakdown of the fluid lubrication layers at different stresses, leading to DST taking place at lower shear rates for higher roughness. In other words, at the same $\phi$, the existence of differences in $\mu$ and $\phi_m$ are hidden until the DST onset and the smaller the value of $\Delta \phi^{*}$, the lower are the observed critical shear rate and shear stress (Figure S5). 
	
	From these results, a clear picture emerges in which the PNIPAM brushes effectively tune the friction, adhesion, and surface roughness of RB\_PNIPAM particles in suspension as a function of temperature. Essentially, the swelling of the brushes can mask surface roughness, reduce friction, and switch off adhesion, so that the DST of the dense suspensions can be significantly retarded. The collapse of the brushes instead introduces adhesive forces, increases friction, and unveils surface roughness so that DST is enhanced. Naturally, changing the temperature across the LCST also causes a shift of the suspension's volume fraction, due to the temperature-dependent particle size. These results have inspired us to utilize the PNIPAM-grafted colloids as a thermo-switchable shear-thickening system, as shown in Figure \ref{Fig4}. Because of the difference in $\phi$, at shear rates of 0.1 s$^{-1}$ and 1 s$^{-1}$, below the DST onset at 40 $^\circ$C, the same dense suspension of RB\_0.52\_PNIPAM exhibits a lower viscosity at 40 $^\circ$C than at 20 $^\circ$C. Nonetheless, the situation is inverted at 10 s$^{-1}$, where the suspension shows DST at 40 $^\circ$C and shear thinning at 20 $^\circ$C. By cycling the temperature up and down during shear, we show that the viscosity switch is reversible because of the reversibility of the swelling-deswelling transition of PNIPAM brushes. 
	
	In conclusion, we have presented a unique approach to use stimuli-responsive polymer brushes, synthesized to a precise length by controlled radical polymerization, for the \textit{in situ}, selective tuning of interparticle friction and adhesion, and for the modulation of surface roughness. Our previous results demonstrated that, in the absence of adhesion, the interlocking of surface asperities promoted dilatant DST  over a broader range of shear rates and for lower values of $\phi$ with increasing (roughness-mediated) effective friction \cite{Hsu_Roughness-dependent_2018}. These experiments additionally show that introducing weak adhesive forces also strengthens ST and shifts it to lower shear rates for smooth contacts. The combination of both effects exacerbates DST for rough adhesive particles. 
	
	This phenomenology presents strong links to recent numerical work, which proposes a constraint-based approach to shear thickening \cite{Guy_Constraint-Based_2018, Singh_shear_2020}. In addition to resisting sliding within shear-induced contact regions as a consequence of friction and topography, adhesion introduces further constraints to the rolling between particles \cite{Dominik_Resistance_1995}, which effectively lowers the high-shear jamming volume fraction, thus promoting ST at a given $\phi$.  
	
	Our results clearly confirm that the microscopic tribology of interparticle contacts is strongly associated with the macroscopic rheology of DST suspensions. This therefore opens up new routes for designing shear-thickening materials via tuning the particle surface chemistry and morphology, in particular developing responsive systems in which friction, adhesion, and surface roughness can be engineered on demand. \\
	
	
	
	
	

	\textbf{Acknowledgements}
	We thank Jan Vermant and Vincent Niggel for fruitful discussions. We thank Jan Vermant for rheometer access, Markus Niederberger for TGA access, and Andre Studart for SEM access. C.-P.H., N.D.S, and L.I. acknowledge financial support from the ETH Research Grant ETH-49 16-1. \\
	
	\textbf{Author Contributions}
	Author contributions are defined based on the CRediT (Contributor Roles Taxonomy) and listed alphabetically. Conceptualization: CPH, LI. Formal analysis: CPH, LI. Funding acquisition: LI, NDS. Investigation: CPH, JM. Methodology: CPH, LI, JM, SNR, NDS. Project administration: LI. Supervision: LI, JM, NDS. Validation: CPH, JM, SNR. Visualization: CPH, LI. Writing – original draft: CPH, LI. Writing – review and editing: CPH, LI, JM, SNR, NDS. \\
	
	
	

	\section{Methods}
	\textbf{Materials.} Silica particles 22 nm (Ludox$^{\circledR}$ TM; DuPont), silica particles 600 nm (Nanocym), aminopropyl-functionalized silica particles 650 nm (Nanocym), tetraethyl orthosilicate (TEOS; 98\%; Sigma-Aldrich), (3-aminopropyl)triethoxysilane (APTES) (97\%; Sigma-Aldrich), $\alpha$-bromoisobutyryl bromide (BIBB) (99\%; Sigma-Aldrich), triethylamine (99.5\%, Sigma-Aldrich), copper(II) bromide (CuBr${_2}$) (99\%; Sigma-Aldrich), tetraethylammonium bromide (TEAB) (98\%; Sigma-Aldrich), tris[2-(dimethylamino)ethyl]amine (Me${_6}$TREN) (97\%; Sigma-Aldrich), polyethyleneimine (PEI) (branched, high molecular weight; Sigma-Aldrich), ethanol (99.8\%, extra dry; ACROS Organics), dichloromethane (DCM) (99.8\%, extra dry; ACROS Organics), silicon wafers (10 mm $\times$ 10 mm; Si-Mat Silicon Wafers) were used as received. N-isopropylacrylamide (NIPAM) (97\%; Sigma-Aldrich) was purified by crystallizing it from a toluene/hexane (3:2, v/v) mixture and was dried under vacuum prior to use. Copper(I) bromide (CuBr) (99.99\%; Sigma-Aldrich) was purified by stirring overnight in glacial acetic acid, filtering, and washing several times with acetone, diethyl ether and was dried under vacuum prior to use.
	
	\textbf{Synthesis of Raspberry-like Silica Particles.} The details of the synthesis of rough particles are described in previous work \cite{Hsu_Roughness-dependent_2018}. In this work, 600-nm and 22-nm silica particles were used as the cores and berries, respectively. Different cycles (6, 9, and 12) of the smoothing process were applied to tune the particles' surface roughness. 
	
	\textbf{Immobilization of SI-ATRP Initiators.} The surfaces of the rough particles were activated and functionalized with APTES in order to graft the initiators for polymerization. 500 mg of rough particles were mixed with 1 mL of APTES in 100 mL of dry ethanol for 12 hours under stirring. The APTES-functionalized rough particles were cleaned three times with ethanol (1055 rcf, 10 min) and dried under vacuum. 500 mg of APTES-functionalized rough particles or aminopropyl-functionalized 650 nm smooth particles were dispersed in 50 mL of dry DCM after degassing with nitrogen for 30 minutes. The functionalized particles were then reacted with 2 mL of BIBB and 4 mL of triethylamine under nitrogen atmosphere for 4 hours. The initiator-grafted particles were cleaned three times (1055 rcf, 10 min) with DCM and dried prior to use.   
	
	\textbf{Synthesis of PNIPAM Brushes on Silica Particles.} SI-ATRP was used to graft the PNIPAM brushes on silica particles. The reaction scheme is illustrated in Figure S1. 300 mg of initiator-grafted particles were dispersed in 13 mL of ethanol/water (4:1, v/v) mixture and degassed with nitrogen for 1 hour. A 25 mL ethanol/water (4:1, v/v) mixture containing 12 g (106 mmol) of NIPAM, 2.04 g (9.7 mmol) of TEAB, and 96 ${\mu}$L (0.35 mmol) of Me${_6}$TREN was degassed with nitrogen for 1 hour and subsequently transferred using a degassed syringe to a flask containing 34.4 mg (0.24 mmol) of CuBr and 26.8 mg (0.12 mmol) of CuBr${_2}$, kept under nitrogen. The solution was then stirred for 10 minutes until complete dissolution of the catalyst. The catalyst-dissolved solution was subsequently transferred using a degassed syringe to the particles solution to carry out the polymerization under nitrogen atmosphere for 30 minutes. The reaction was quenched by adding 40 mL of ethanol/water (4:1, v/v) mixture and by exposing the reaction mixture to air. The PNIPAM-grafted particles were cleaned three times with Milli-Q water (1055 rcf, 10 min), and kept in Milli-Q water for further use.
	
	\textbf{Characterization of PNIPAM Brushes on Silica Particles.} Dynamic light scattering (DLS) (Zetasizer Nano ZS; Malvern Panalytical) experiments were performed at 20 $^\circ$C and 40 $^\circ$C in Miilli-Q water to determine the particles' size. The thickness of PNIPAM brushes $h_{\text{PNIPAM}}$ were obtained by comparing the particles' hydrodynamic radii before ($R_{\text{h}_0}$) and after ($R_\text{h}$) the polymerization. Thermogravimetric analysis (TGA) (TGA/DSC 3+; METTLER TOLEDO) was used to determine the mass fraction of PNIPAM brushes $m_{\text{PNIPAM}}$ and silica particles $m_{\text{SiO}_2}$ of the PNIPAM-grafted particles which $m_{\text{PNIPAM}}/m_{\text{SiO}_2} = (m_{850 ^{\circ}C} - m_{250 ^{\circ}C})/m_{850 ^{\circ}C}$ (Figure S6). The sample was heated up to 80 $^\circ$C at 10 $^\circ$C/min and kept at 80 $^\circ$C for one hour to remove the water. The sample was then heated up to 900 $^\circ$C at 10 $^\circ$C/min.
	
	\textbf{Fabrication of Rough Substrates.} The particle-coated silicon wafers were fabricated according to previous work \cite{Hsu_Roughness-dependent_2018}. PEI-coated silicon wafers were immersed in 0.004 wt\% of 22-nm silica particle suspensions for 20 minutes. Different surface roughness were achieved by adding different amount (0.6, 0.9, and 1.2 mL) of 1 vol\% TEOS in ethanol solution in the smoothing process.
	
	\textbf{Synthesis of PNIPAM Brushes on Planar Substrates.} The same protocol used for particles was applied to graft the PNIPAM brushes on smooth and rough substrates. The substrates ($\sim$12--15 pieces) were functionalized with APTES in dry ethanol and subsequently rinsed with ethanol and blown dry with a nitrogen jet. The APTES-modified substrates were then reacted with BIBB and triethylamine in dry DCM after degassing under nitrogen. The reaction was continued for 4 hours and finally the substrates were cleaned with DCM and blown dry with a nitrogen jet. Instead of 300 mg of particles, 6 pieces of smooth or rough substrates were used in the SI-ATRP process. The reaction time for the smooth and rough substrates was 20 and 25 minutes, respectively. The difference in reaction time was to achieve PNIPAM brushes with thicknesses close to the PNIPAM-grafted smooth and rough particles. The PNIPAM-grafted substrates were cleaned with toluene and blown dry with a nitrogen jet. Smooth PNIPAM-grafted substrates are named “SM\_PNIPAM” and rough PNIPAM-grafted substrates are named “RS\_$h/d$\_PNIPAM”, where $h/d$ is the value of the dimensionless roughness parameter of the silica surface measured before polymerization. The SEM images are shown in Figure S2. 
	
	\textbf{Characterization of PNIPAM Brushes on Planar Substrates.} The thickness of the PNIPAM brushes was measured using variable-angle spectroscopic ellipsometry (VASE; M-2000F; J. A. Woollam Co.). All data were recorded between 275 and 827 nm of wavelength at 70$^\circ$ with respect to the surface. The obtained values of amplitude ($\Psi$) and phase ($\Delta$) components were analyzed with WVASE32 software using a three-layer model (Si/SiO$_2$/Cauchy; $A_n$ = 1.45, $B_n$ = 0.01, and $C_n$ = 0). All measurements were performed in Milli-Q water at 20 $^\circ$C and 40 $^\circ$C. The thickness of the rough SiO$_2$ layer on the substrate was measured before the polymerization and the value was used in the three-layer model to determine the PNIPAM thickness. The measured PNIPAM thickness are listed in Table SI.
	
	\textbf{Centrifugation Experiments.} The centrifugation experiments were performed based on previous work \cite{Hsu_Roughness-dependent_2018}. The experiments were carried out at 20 $^\circ$C and 40 $^\circ$C using a temperature-controlled centrifuge (Avanti$\circledR$ J-25I; Beckman Coulter). The effective volume fraction $\phi_{\text{eff}}$ of each sample was calculated in three steps: 1) the wt\% of silica particles was obtained by $ wt\% (\text{SiO}_2) = wt\% (\text{total}) \cdot (m_{\text{SiO}_2}/m_{\text{PNIPAM}} + m_{\text{SiO}_2})$, according to the TGA measurements. 2) $wt\% (\text{SiO}_2)$ was converted into the volume fraction of silica particles $\phi_{\text{SiO}_2}$ using a silica density of 1.8 g/cm$^3$ \cite{Bogush_Preparation_1988}. 3) $\phi_{\text{eff}} = \phi_{\text{SiO}_2} \cdot (R_\text{h}(\text{T})/R_{\text{h}_0}(\text{T}))^3$ was calculated based on the DLS measurements. The frictional packing fraction $\phi_m$ was evaluated after the centrifugation and the maximal packing fraction $\phi_f$ was determined after ten days for suspensions kept at 20 $^\circ$C and 40 $^\circ$C, respectively (Figure S7).
	
	\textbf{Shear Rheology Measurements.} The shear rheology measurements were performed on a rheometer (MCR 302; Anton Paar) at 20 $^\circ$C and 40 $^\circ$C in a cone-and-plate geometry, using a 25 mm diameter stainless steel cone with an angle of 2$^\circ$. The flow curves were recorded for increasing shear rates from 0.1 to 100 s$^{-1}$. The suspensions were prepared by centrifugation (2701 rcf, 15 min) to produce the sediments at $\phi_m$. Different amounts of Milli-Q water were added to the different sediments to adjust $\phi$ to the desired volume fraction. The suspensions were then sonicated for 30 min before performing the measurements. The viscosity at fixed shear rate (0.1, 1, and 10 s$^{-1}$) was measured alternately between 20 $^\circ$C and 40 $^\circ$C. In all cases, in order to reset the normal force before measuring, the samples were rejuvenated by applying 1\% shear strain oscillation for 1 minute.   
	
	\textbf{Adhesion and Friction Measurements.} Atomic force microscopy (AFM) (NanoWizard$\circledR$ NanoOptics; JPK Instruments AG) was used to measure the adhesion and friction forces between a PNIPAM-grafted particle and a PNIPAM-grafted substrate. All the measurements were carried out at 20 $^\circ$C and 40 $^\circ$C in a liquid cell (BiOCell$^{TM}$; JPK Instruments AG). The adhesion force was determined by measuring the pull-off force in a force versus distance curve. 200 curves were recorded across a 20 $\mu$ m $\times$ 20 $\mu$m area with the approach and retraction speed set at 500 nm/s. The preparation of colloidal probes and the details of friction measurements are described in previous work \cite{Hsu_Roughness-dependent_2018}. In this work, we used micromanipulator miBot (Imina Technologies SA) to affix the colloidal particles onto the tipless cantilevers. For the friction measurements, the scan area and scan rate were fixed at 2 $\mu$m $\times$ 200 nm (512 px $\times$ 51 px) and 1 Hz, respectively. At every scanned area, friction loops were recorded at different applied loads $L$ from 10 nN to 60 nN. The average friction force $F_{\text{friction}}$ for a given load was obtained by averaging half the vertical difference between the trace and retrace curves for all the friction loops.

    \bibliographystyle{naturemag}
	\bibliography{ref}

\begin{thebibliography}{10}
\expandafter\ifx\csname url\endcsname\relax
  \def\url#1{\texttt{#1}}\fi
\expandafter\ifx\csname urlprefix\endcsname\relax\def\urlprefix{URL }\fi
\providecommand{\bibinfo}[2]{#2}
\providecommand{\eprint}[2][]{\url{#2}}

\bibitem{Brown_Shear_2014}
\bibinfo{author}{Brown, E.} \& \bibinfo{author}{Jaeger, H.~M.}
\newblock \bibinfo{title}{Shear thickening in concentrated suspensions:
  phenomenology, mechanisms and relations to jamming}.
\newblock \emph{\bibinfo{journal}{Rep Prog Phys}}
  \textbf{\bibinfo{volume}{77}}, \bibinfo{pages}{046602}
  (\bibinfo{year}{2014}).

\bibitem{Wagner_Shear_2009}
\bibinfo{author}{Wagner, N.~J.} \& \bibinfo{author}{Brady, J.~F.}
\newblock \bibinfo{title}{Shear thickening in colloidal dispersions}.
\newblock \emph{\bibinfo{journal}{Physics Today}}
  \textbf{\bibinfo{volume}{62}}, \bibinfo{pages}{27--32}
  (\bibinfo{year}{2009}).

\bibitem{Brown_Generality_2010}
\bibinfo{author}{Brown, E.} \emph{et~al.}
\newblock \bibinfo{title}{Generality of shear thickening in dense suspensions}.
\newblock \emph{\bibinfo{journal}{Nat Mater}} \textbf{\bibinfo{volume}{9}},
  \bibinfo{pages}{220--224} (\bibinfo{year}{2010}).

\bibitem{Seto_Discontinous_2013}
\bibinfo{author}{Seto, R.}, \bibinfo{author}{Mari, R.},
  \bibinfo{author}{Morris, J.~F.} \& \bibinfo{author}{Denn, M.~M.}
\newblock \bibinfo{title}{Discontinuous shear thickening of frictional
  hard-sphere suspensions}.
\newblock \emph{\bibinfo{journal}{Physical Review Letters}}
  \textbf{\bibinfo{volume}{111}} (\bibinfo{year}{2013}).

\bibitem{Fernandez_Microscopic_2013}
\bibinfo{author}{Fernandez, N.} \emph{et~al.}
\newblock \bibinfo{title}{Microscopic mechanism for shear thickening of
  non-brownian suspensions}.
\newblock \emph{\bibinfo{journal}{Physical Review Letters}}
  \textbf{\bibinfo{volume}{111}}, \bibinfo{pages}{108301}
  (\bibinfo{year}{2013}).

\bibitem{Wyart_Discontinuous_2014}
\bibinfo{author}{Wyart, M.} \& \bibinfo{author}{Cates, M.~E.}
\newblock \bibinfo{title}{Discontinuous shear thickening without inertia in
  dense non-brownian suspensions}.
\newblock \emph{\bibinfo{journal}{Phys Rev Lett}}
  \textbf{\bibinfo{volume}{112}}, \bibinfo{pages}{098302}
  (\bibinfo{year}{2014}).

\bibitem{Cates_Jamming_1998}
\bibinfo{author}{Cates, M.~E.}, \bibinfo{author}{Wittmer, J.~P.},
  \bibinfo{author}{Bouchaud, J.~P.} \& \bibinfo{author}{Claudin, P.}
\newblock \bibinfo{title}{Jamming, force chains, and fragile matter}.
\newblock \emph{\bibinfo{journal}{Physical Review Letters}}
  \textbf{\bibinfo{volume}{81}}, \bibinfo{pages}{1841--1844}
  (\bibinfo{year}{1998}).

\bibitem{Waitukaitis_Impact-activated_2012}
\bibinfo{author}{Waitukaitis, S.~R.} \& \bibinfo{author}{Jaeger, H.~M.}
\newblock \bibinfo{title}{Impact-activated solidification of dense suspensions
  via dynamic jamming fronts}.
\newblock \emph{\bibinfo{journal}{Nature}} \textbf{\bibinfo{volume}{487}},
  \bibinfo{pages}{205--209} (\bibinfo{year}{2012}).

\bibitem{Peters_Direct_2016}
\bibinfo{author}{Peters, I.~R.}, \bibinfo{author}{Majumdar, S.} \&
  \bibinfo{author}{Jaeger, H.~M.}
\newblock \bibinfo{title}{Direct observation of dynamic shear jamming in dense
  suspensions}.
\newblock \emph{\bibinfo{journal}{Nature}} \textbf{\bibinfo{volume}{532}},
  \bibinfo{pages}{214--217} (\bibinfo{year}{2016}).

\bibitem{Lin_Hydrodynamic_2015}
\bibinfo{author}{Lin, N. Y.~C.} \emph{et~al.}
\newblock \bibinfo{title}{Hydrodynamic and contact contributions to continuous
  shear thickening in colloidal suspensions}.
\newblock \emph{\bibinfo{journal}{Physical Review Letters}}
  \textbf{\bibinfo{volume}{115}} (\bibinfo{year}{2015}).

\bibitem{Royer_Rheological_2016}
\bibinfo{author}{Royer, J.~R.}, \bibinfo{author}{Blair, D.~L.} \&
  \bibinfo{author}{Hudson, S.~D.}
\newblock \bibinfo{title}{Rheological signature of frictional interactions in
  shear thickening suspensions}.
\newblock \emph{\bibinfo{journal}{Physical Review Letters}}
  \textbf{\bibinfo{volume}{116}} (\bibinfo{year}{2016}).

\bibitem{Clavaud_Revealing_2017}
\bibinfo{author}{Clavaud, C.}, \bibinfo{author}{Berut, A.},
  \bibinfo{author}{Metzger, B.} \& \bibinfo{author}{Forterre, Y.}
\newblock \bibinfo{title}{Revealing the frictional transition in
  shear-thickening suspensions}.
\newblock \emph{\bibinfo{journal}{Proceedings of the National Academy of
  Sciences of the United States of America}} \textbf{\bibinfo{volume}{114}},
  \bibinfo{pages}{5147--5152} (\bibinfo{year}{2017}).

\bibitem{Morris_Lubricated_2018}
\bibinfo{author}{Morris, J.~F.}
\newblock \bibinfo{title}{Lubricated-to-frictional shear thickening scenario in
  dense suspensions}.
\newblock \emph{\bibinfo{journal}{Physical Review Fluids}}
  \textbf{\bibinfo{volume}{3}} (\bibinfo{year}{2018}).

\bibitem{Singh_From_2019}
\bibinfo{author}{Singh, A.}, \bibinfo{author}{Pednekar, S.},
  \bibinfo{author}{Chun, J.}, \bibinfo{author}{Denn, M.~M.} \&
  \bibinfo{author}{Morris, J.~F.}
\newblock \bibinfo{title}{From yielding to shear jamming in a cohesive
  frictional suspension}.
\newblock \emph{\bibinfo{journal}{Physical Review Letters}}
  \textbf{\bibinfo{volume}{122}} (\bibinfo{year}{2019}).

\bibitem{Lootens_Dilatant_2005}
\bibinfo{author}{Lootens, D.}, \bibinfo{author}{van Damme, H.},
  \bibinfo{author}{Hemar, Y.} \& \bibinfo{author}{Hebraud, P.}
\newblock \bibinfo{title}{Dilatant flow of concentrated suspensions of rough
  particles}.
\newblock \emph{\bibinfo{journal}{Phys Rev Lett}}
  \textbf{\bibinfo{volume}{95}}, \bibinfo{pages}{268302}
  (\bibinfo{year}{2005}).

\bibitem{Hsiao_Rheological_2017}
\bibinfo{author}{Hsiao, L.~C.} \emph{et~al.}
\newblock \bibinfo{title}{Rheological state diagrams for rough colloids in
  shear flow}.
\newblock \emph{\bibinfo{journal}{Physical Review Letters}}
  \textbf{\bibinfo{volume}{119}}, \bibinfo{pages}{158001}
  (\bibinfo{year}{2017}).

\bibitem{Hsu_Roughness-dependent_2018}
\bibinfo{author}{Hsu, C.~P.}, \bibinfo{author}{Ramakrishna, S.~N.},
  \bibinfo{author}{Zanini, M.}, \bibinfo{author}{Spencer, N.~D.} \&
  \bibinfo{author}{Isa, L.}
\newblock \bibinfo{title}{Roughness-dependent tribology effects on
  discontinuous shear thickening}.
\newblock \emph{\bibinfo{journal}{Proc Natl Acad Sci U S A}}
  \textbf{\bibinfo{volume}{115}}, \bibinfo{pages}{5117--5122}
  (\bibinfo{year}{2018}).

\bibitem{James_Interparticle_2018}
\bibinfo{author}{James, N.~M.}, \bibinfo{author}{Han, E.},
  \bibinfo{author}{de~la Cruz, R. A.~L.}, \bibinfo{author}{Jureller, J.} \&
  \bibinfo{author}{Jaeger, H.~M.}
\newblock \bibinfo{title}{Interparticle hydrogen bonding can elicit shear
  jamming in dense suspensions}.
\newblock \emph{\bibinfo{journal}{Nat Mater}} \textbf{\bibinfo{volume}{17}},
  \bibinfo{pages}{965--970} (\bibinfo{year}{2018}).

\bibitem{James_Tuning_2019}
\bibinfo{author}{James, N.~M.}, \bibinfo{author}{Hsu, C.~P.},
  \bibinfo{author}{Spencer, N.~D.}, \bibinfo{author}{Jaeger, H.~M.} \&
  \bibinfo{author}{Isa, L.}
\newblock \bibinfo{title}{Tuning interparticle hydrogen bonding in
  shear-jamming suspensions: Kinetic effects and consequences for tribology and
  rheology}.
\newblock \emph{\bibinfo{journal}{Journal of Physical Chemistry Letters}}
  \textbf{\bibinfo{volume}{10}}, \bibinfo{pages}{1663--1668}
  (\bibinfo{year}{2019}).

\bibitem{Matyjaszewski_Polymers_1999}
\bibinfo{author}{Matyjaszewski, K.} \emph{et~al.}
\newblock \bibinfo{title}{Polymers at interfaces: Using atom transfer radical
  polymerization in the controlled growth of homopolymers and block copolymers
  from silicon surfaces in the absence of untethered sacrificial initiator}.
\newblock \emph{\bibinfo{journal}{Macromolecules}}
  \textbf{\bibinfo{volume}{32}}, \bibinfo{pages}{8716--8724}
  (\bibinfo{year}{1999}).

\bibitem{Mandal_Tuning_2019}
\bibinfo{author}{Mandal, J.}, \bibinfo{author}{Simic, R.} \&
  \bibinfo{author}{Spencer, N.~D.}
\newblock \bibinfo{title}{Tuning and in situ monitoring of surface-initiated,
  atom-transfer radical polymerization of acrylamide derivatives in water-based
  solvents}.
\newblock \emph{\bibinfo{journal}{Polymer Chemistry}}
  \textbf{\bibinfo{volume}{10}}, \bibinfo{pages}{3933--3942}
  (\bibinfo{year}{2019}).

\bibitem{Halperin_Poly_2015}
\bibinfo{author}{Halperin, A.}, \bibinfo{author}{Kroger, M.} \&
  \bibinfo{author}{Winnik, F.~M.}
\newblock \bibinfo{title}{Poly(n-isopropylacrylamide) phase diagrams: Fifty
  years of research}.
\newblock \emph{\bibinfo{journal}{Angewandte Chemie-International Edition}}
  \textbf{\bibinfo{volume}{54}}, \bibinfo{pages}{15342--15367}
  (\bibinfo{year}{2015}).

\bibitem{Yu_Stretching_2015}
\bibinfo{author}{Yu, Y.} \emph{et~al.}
\newblock \bibinfo{title}{Stretching of collapsed polymers causes an enhanced
  dissipative response of pnipam brushes near their lcst}.
\newblock \emph{\bibinfo{journal}{Soft Matter}} \textbf{\bibinfo{volume}{11}},
  \bibinfo{pages}{8508--8516} (\bibinfo{year}{2015}).

\bibitem{Chang_Switchable_2007}
\bibinfo{author}{Chang, D.~P.}, \bibinfo{author}{Dolbow, J.~E.} \&
  \bibinfo{author}{Zauscher, S.}
\newblock \bibinfo{title}{Switchable friction of stimulus-responsive
  hydrogels}.
\newblock \emph{\bibinfo{journal}{Langmuir}} \textbf{\bibinfo{volume}{23}},
  \bibinfo{pages}{250--257} (\bibinfo{year}{2007}).

\bibitem{Jones_Variable_2002}
\bibinfo{author}{Jones, D.~M.}, \bibinfo{author}{Smith, J.~R.},
  \bibinfo{author}{Huck, W. T.~S.} \& \bibinfo{author}{Alexander, C.}
\newblock \bibinfo{title}{Variable adhesion of micropatterned thermoresponsive
  polymer brushes: Afm investigations of poly (n-isopropylacrylamide) brushes
  prepared by surface-initiated polymerizations}.
\newblock \emph{\bibinfo{journal}{Advanced Materials}}
  \textbf{\bibinfo{volume}{14}}, \bibinfo{pages}{1130--1134}
  (\bibinfo{year}{2002}).

\bibitem{Divandari_Topology_2017}
\bibinfo{author}{Divandari, M.} \emph{et~al.}
\newblock \bibinfo{title}{Topology effects on the structural and
  physicochemical properties of polymer brushes}.
\newblock \emph{\bibinfo{journal}{Macromolecules}}
  \textbf{\bibinfo{volume}{50}}, \bibinfo{pages}{7760--7769}
  (\bibinfo{year}{2017}).

\bibitem{Derjaguin_Molecular_1934}
\bibinfo{author}{Derjaguin, B.}
\newblock \bibinfo{title}{Molecular theory of outer friction}.
\newblock \emph{\bibinfo{journal}{Zeitschrift Fur Physik}}
  \textbf{\bibinfo{volume}{88}}, \bibinfo{pages}{661--675}
  (\bibinfo{year}{1934}).

\bibitem{Luo_One_2020}
\bibinfo{author}{Luo, Y.} \emph{et~al.}
\newblock \bibinfo{title}{One-step, in situ jamming point measurements by
  immobilization cell rheometry}.
\newblock \emph{\bibinfo{journal}{Rheologica Acta}}  (\bibinfo{year}{2020}).

\bibitem{Guy_Constraint-Based_2018}
\bibinfo{author}{Guy, B.~M.}, \bibinfo{author}{Richards, J.~A.},
  \bibinfo{author}{Hodgson, D. J.~M.}, \bibinfo{author}{Blanco, E.} \&
  \bibinfo{author}{Poon, W. C.~K.}
\newblock \bibinfo{title}{Constraint-based approach to granular dispersion
  rheology}.
\newblock \emph{\bibinfo{journal}{Physical Review Letters}}
  \textbf{\bibinfo{volume}{121}} (\bibinfo{year}{2018}).

\bibitem{Singh_shear_2020}
\bibinfo{author}{Singh, A.}, \bibinfo{author}{Ness, C.}, \bibinfo{author}{Seto,
  R.}, \bibinfo{author}{de~Pablo, J.~J.} \& \bibinfo{author}{Jaeger, H.~M.}
\newblock \bibinfo{title}{Shear thickening and jamming of dense suspensions:
  the roll of friction} (\bibinfo{year}{2020}).
\newblock \eprint{arXiv:2002.10996}.

\bibitem{Dominik_Resistance_1995}
\bibinfo{author}{Dominik, C.} \& \bibinfo{author}{Tielens, A. G. G.~M.}
\newblock \bibinfo{title}{Resistance to rolling in the adhesive contact of two
  elastic spheres}.
\newblock \emph{\bibinfo{journal}{Philosophical Magazine A}}
  \textbf{\bibinfo{volume}{72}}, \bibinfo{pages}{783--803}
  (\bibinfo{year}{1995}).

\bibitem{Bogush_Preparation_1988}
\bibinfo{author}{Bogush, G.~H.}, \bibinfo{author}{Tracy, M.~A.} \&
  \bibinfo{author}{Zukoski, C.~F.}
\newblock \bibinfo{title}{Preparation of monodisperse silica particles -
  control of size and mass fraction}.
\newblock \emph{\bibinfo{journal}{Journal of Non-Crystalline Solids}}
  \textbf{\bibinfo{volume}{104}}, \bibinfo{pages}{95--106}
  (\bibinfo{year}{1988}).

\end{thebibliography}
	
\end{document}


\title{Supplementary Information for \\ Disentangling the roles of roughness, friction and adhesion in discontinuous shear thickening by means of thermo-responsive particles}
	\author{Chiao-Peng Hsu}
	\affiliation{Laboratory for Soft Materials and Interfaces, Department of Materials, ETH Zurich, Zurich, Switzerland.}
	\affiliation{Laboratory for Surface Science and Technology, Department of Materials, ETH Zurich, Zurich, Switzerland.}
	\author{Joydeb Mandal}
	\affiliation{Laboratory for Surface Science and Technology, Department of Materials, ETH Zurich, Zurich, Switzerland.}
	\author{Shivaprakash N. Ramakrishna}
	\affiliation{Laboratory for Surface Science and Technology, Department of Materials, ETH Zurich, Zurich, Switzerland.}
	\author{Nicholas D. Spencer}
	\affiliation{Laboratory for Surface Science and Technology, Department of Materials, ETH Zurich, Zurich, Switzerland.}
	\author{Lucio Isa}
	\affiliation{Laboratory for Soft Materials and Interfaces, Department of Materials, ETH Zurich, Zurich, Switzerland.}
	\maketitle
	
	\begin{figure*}
		\centering
		\includegraphics[width=0.95\textwidth]{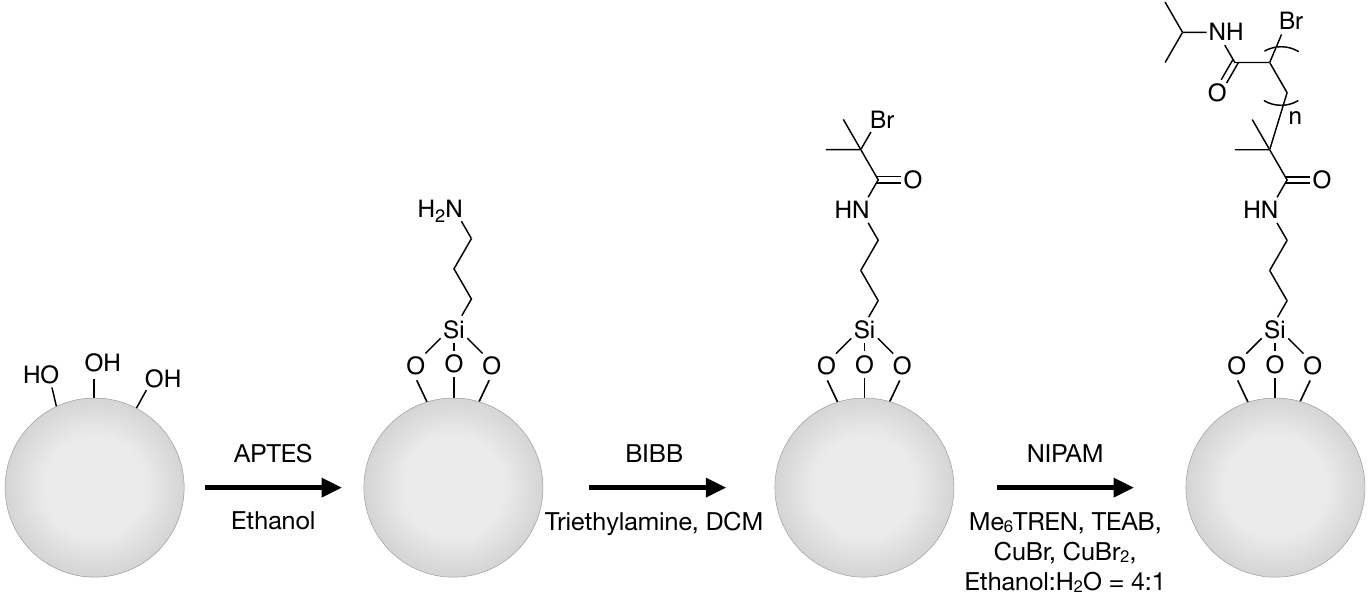}
		\caption{Reaction scheme to prepare PNIPAM brushes on the surface of a silica particle using SI-ATRP.}
		\label{FigS1}
	\end{figure*}

	\begin{figure*}
		\centering
		\includegraphics[width=0.6\textwidth]{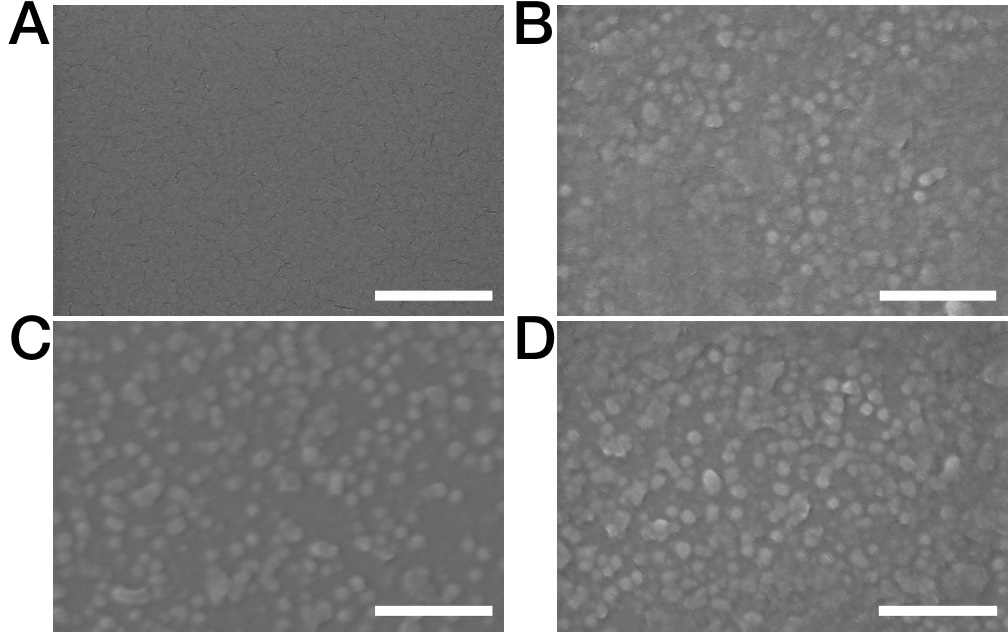}
		\caption{\textbf{PNIPAM-grafted Substrates.} SEM images of SS\_PNIPAM (\textbf{a}), RS\_0.36\_PNIPAM (\textbf{b}), RS\_0.46\_PNIPAM (\textbf{c}), RS\_0.52\_PNIPAM (\textbf{d}). Scale bars indicate 200 nm. }
		\label{FigS2}
	\end{figure*} 
	
	\begin{figure*}
		\centering
		\includegraphics[width=0.4\textwidth]{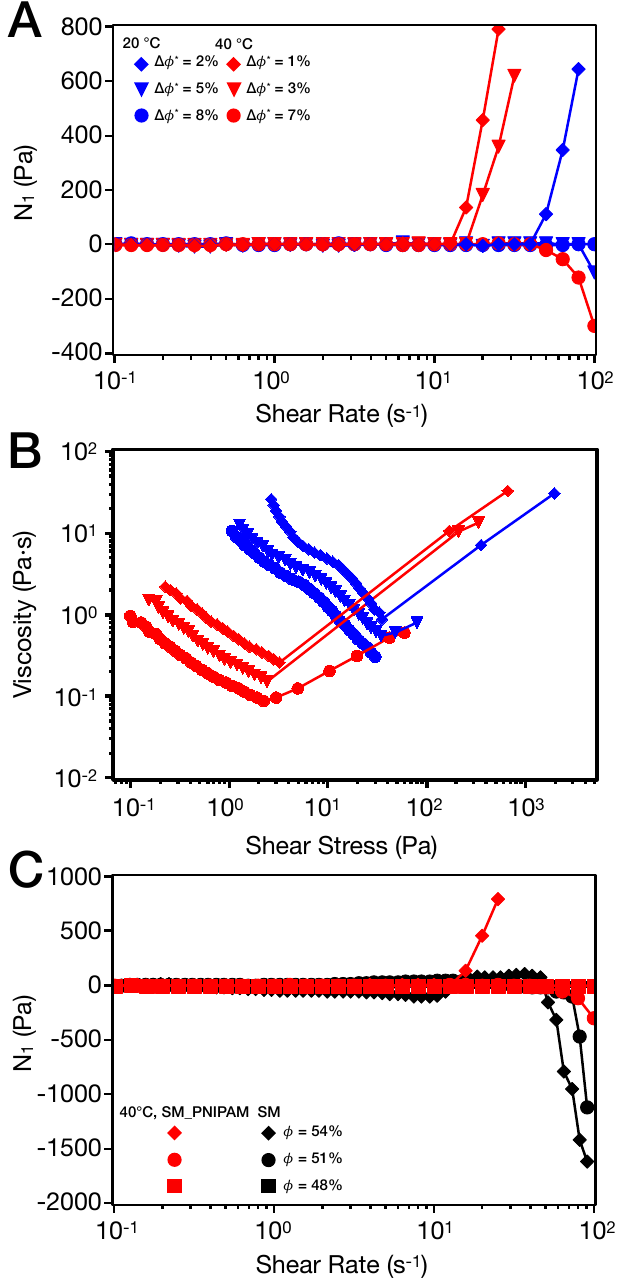}
		\caption{\textbf{Shear-rheology experiments of PNIPAM-grafted smooth particles.} \textbf{a}, N$_1$ of the SM\_PNIPAM flow curves shown in Figure 2e with the same legends. \textbf{b}, Flow curves from Figure 2e plotted as viscosity vs. stress with the same legends. \textbf(c), N$_1$ of the normalized flow curves shown in Figure 2f with the same legends.}
		\label{Fig3}
	\end{figure*} 
	
	\begin{figure*}
		\centering
		\includegraphics[width=0.7\textwidth]{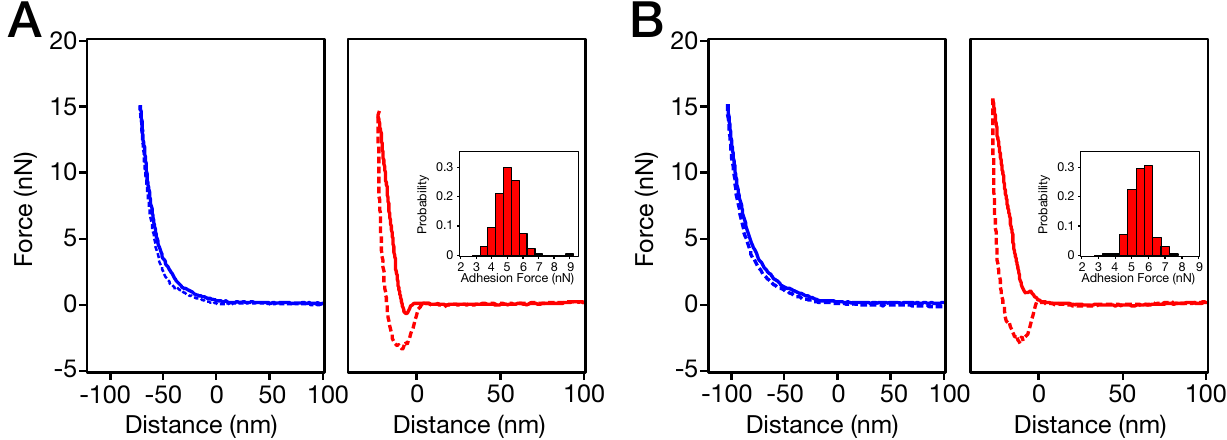}
		\caption{\textbf{Force--distance curves of PNIPAM-grafted rough particles.} \textbf{a}, Approach (solid line) and retraction (dashed line) of the force--distance curves of the RB\_0.46\_PNIPAM system at 20 $^\circ$C (left) and 40 $^\circ$C (right). The inset shows the adhesion forces (5.0 $\pm$ 1.3 nN) measured at 40 $^\circ$C. The bin width is 0.5 nN. \textbf{b}, Approach (solid line) and retraction (dashed line) of the force--distance curves of the RB\_0.36\_PNIPAM system at 20 $^\circ$C (left) and 40 $^\circ$C (right). The inset shows the adhesion forces (5.5 $\pm$ 1.2 nN) measured at 40 $^\circ$C. The bin width is 0.5 nN.}
		\label{FigS4}
	\end{figure*}
	
	\begin{figure*}
		\centering
		\includegraphics[width=0.7\textwidth]{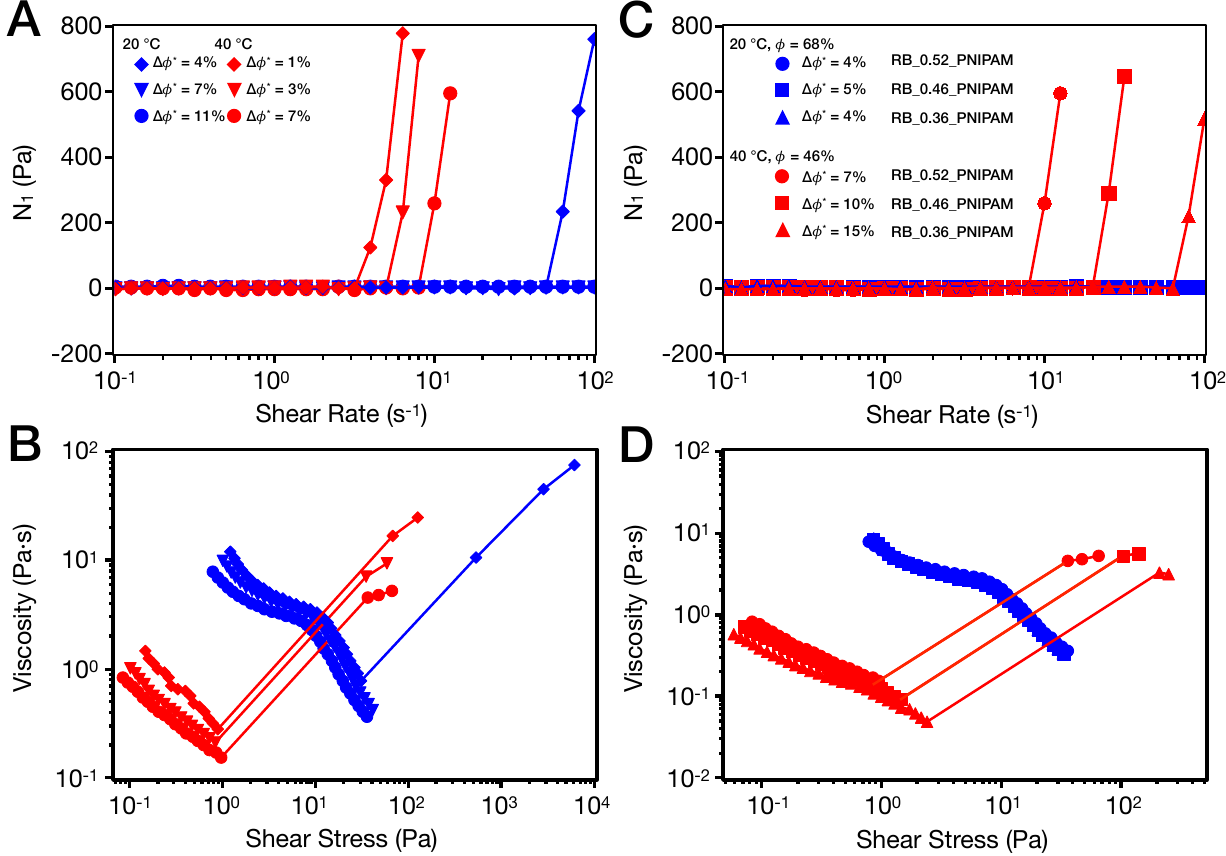}
		\caption{\textbf{Shear-rheology experiments of PNIPAM-grafted rough particles.} \textbf{a}, N$_1$ of the RB\_0.52\_PNIPAM flow curves shown in Figure 3e with the same legends. \textbf{b}, Flow curves from Figure 3e plotted as viscosity vs. stress with the same legends. \textbf{c}, N$_1$ of the three PNIPAM-grafted rough systems shown in Figure 3f with the same legends. \textbf{d}, Flow curves from Figure 3f plotted as viscosity vs. stress with the same legends.}
		\label{FigS5}
	\end{figure*} 
	
	\begin{figure*}
		\centering
		\includegraphics[width=0.5\textwidth]{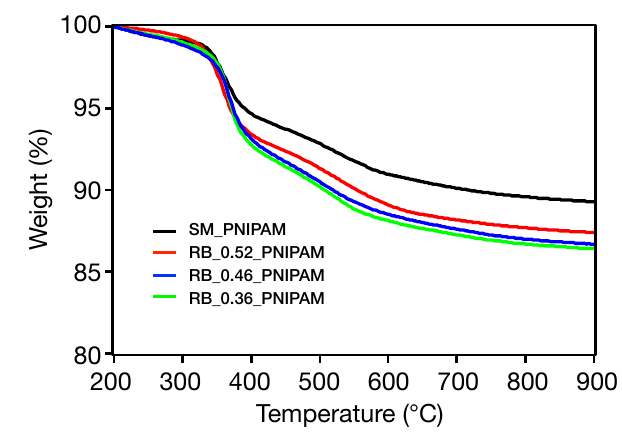}
		\caption{\textbf{Determination of PNIPAM brush content.} TGA curves of SM\_PNIPAM, black; RB\_0.52\_PNIPAM, red; RB\_0.46\_PNIPAM, blue; RB\_0.36\_PNIPAM; green.}
		\label{Fig6}
	\end{figure*}
	
	\begin{figure*}
		\centering
		\includegraphics[width=0.8\textwidth]{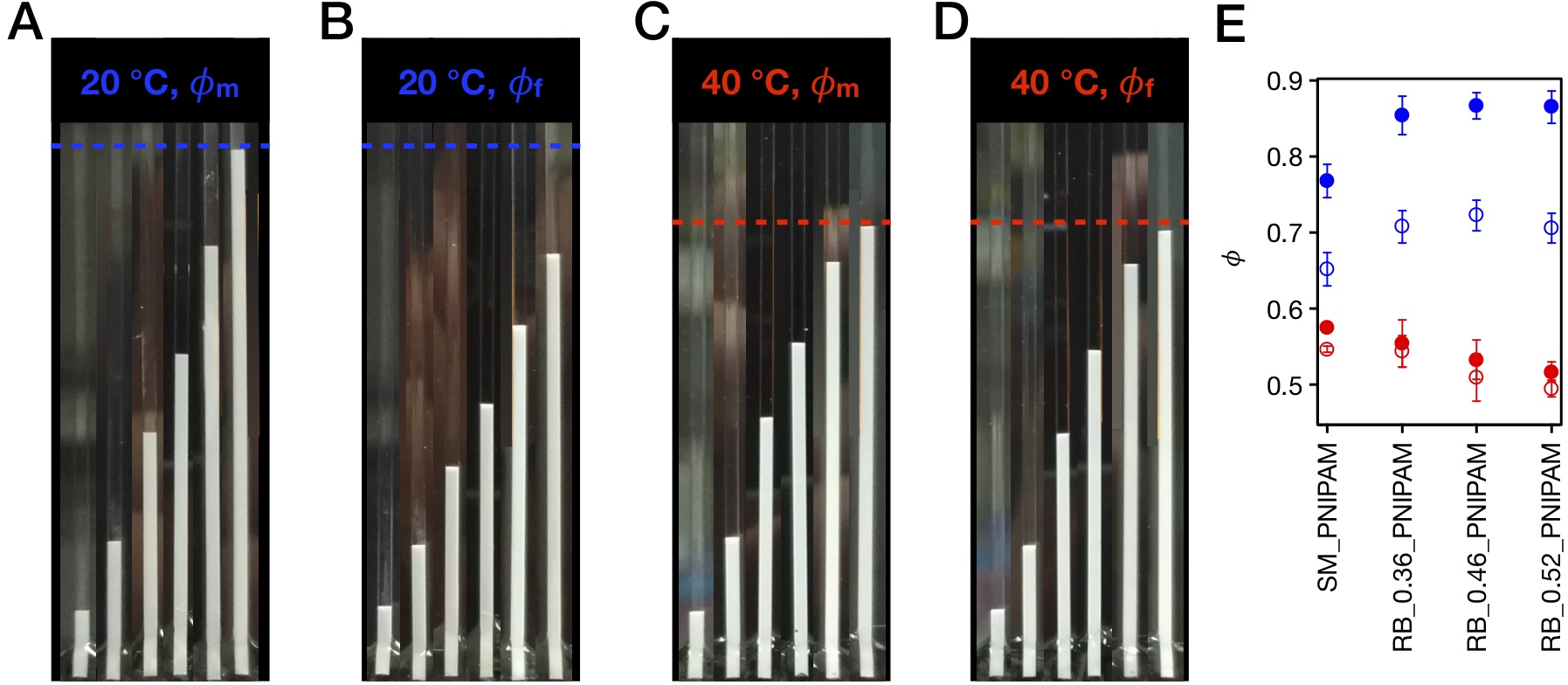}
		\caption{\textbf{Centrifugation experiments and packing fractions.} Images of a particle suspension (RB\_0.52\_PNIPAM) (\textbf{a}), immediately after centrifugation at 20 $^\circ$C for determination of $\phi_m$, (\textbf{b}), 10 days after centrifugation at 20 $^\circ$ for determination of $\phi_f$, (\textbf{c}), immediately after centrifugation at 40 $^\circ$C for determination of $\phi_m$, (\textbf{d}) 10 days after centrifugation at 40 $^\circ$ for determination of $\phi_m$. The initial volume fraction $\phi_i$ increases from 9.7\% (left) to 57.6\% (right) for (\textbf{a}) and (\textbf{b}), and from 7.2\% (left) to 42.3\% (right) for (\textbf{c}) and (\textbf{d}). The blue dotted line in (\textbf{a}) and (\textbf{b}) marks the height of $\phi_i$ = 57.6\% immediately after centrifugation. The red dotted line in (\textbf{c}) and (\textbf{d}) marks the height of $\phi_i$ = 42.3\% immediately after centrifugation. (\textbf{e}), $\phi_{m}$ (open) and $\phi_{f}$ (filled) of the four colloidal suspensions at 20 $^\circ$C (blue) and 40 $^\circ$C (red).}
		\label{FigS7}
	\end{figure*}
	
	\begin{table*}[htbp]
		\caption{\textbf{PNIPAM Thickness of the PNIPAM-grafted Substrates.} $h_{\text{PNIPAM}}(20 ^{\circ}C)$ is the swollen PNIPAM thickness and $h_{\text{PNIPAM}}(40 ^{\circ}C)$ is the collapsed PNIPAM thickness.}
		\centering
		\begin{tabular}{ccccc}		
			\hline 
			&SS\_PNIPAM &RS\_0.36\_PNIPAM &RS\_0.46\_PNIPAM &RS\_0.52\_PNIPAM  \\ 
			\hline   
			$h_{\text{PNIPAM}}(20 ^{\circ}C)$ (nm) &35 $\pm$ 2 &83 $\pm$ 5 &88 $\pm$ 4 &85 $\pm$ 3 \\
			$h_{\text{PNIPAM}}(40 ^{\circ}C)$ (nm) &14 $\pm$ 1 &34 $\pm$ 2 &35 $\pm$ 3 &31 $\pm$ 2 \\
			\hline
		\end{tabular}
		\label{Tab1}
	\end{table*}